\documentclass[usenatbib]{mnras}
\usepackage[utf8]{inputenc}
\usepackage{aas_macros}
\usepackage{amsfonts}
\usepackage{amsmath}
\usepackage{bm}
\usepackage{amssymb}
\usepackage{graphicx}
\usepackage[dvipsnames]{xcolor}
\usepackage{hyperref}
\hypersetup{colorlinks=true,allcolors=teal}
\usepackage{microtype}

\defcitealias{ps21}{PS21}
\defcitealias{pcs21}{PCS21}
\defcitealias{efe20}{C20}
\defcitealias{dflj18}{D18}

\newcommand{\Hi}{\textsc{Hi}}

\newcommand{\atot}{\ensuremath{a_{\rm tot}}}
\newcommand{\abary}{\ensuremath{a_{\rm bary}}}
\newcommand{\aext}{\ensuremath{\mathbf{a}_{\rm ext}}}

\newcommand{\Msun}{\ensuremath{M_{\odot}}}
\newcommand{\Mh}{\ensuremath{h^{-1}M_{\odot}}}
\newcommand{\Mhsq}{\ensuremath{h^{-2}M_{\odot}}}
\newcommand{\Mpch}{\ensuremath{h^{-1}{\rm Mpc}}}

\newcommand{\msq}{\ensuremath{{\rm \,m\,s}^{-2}}}

\newcommand{\avg}[1]{\ensuremath{\left\langle \,#1\, \right\rangle}}
\newcommand{\e}[1]{\ensuremath{{\rm e}^{#1}}}

\newcommand{\der}{\ensuremath{{\rm d}}}

\newcommand{\eqn}[1]{equation~\eqref{#1}}

\newcommand{\be}{\begin{equation}}
\newcommand{\ee}{\end{equation}}
\newcommand{\Cal}[1]{\ensuremath{\mathcal{#1}}}

\title[EFE in CDM]{The phenomenology of the external field effect in cold dark matter models} 
\author[Paranjape \& Sheth]{
 Aseem Paranjape$^{1}$\thanks{E-mail: aseem@iucaa.in} \& 
 Ravi K. Sheth$^{2,3}$\thanks{E-mail: shethrk@physics.upenn.edu},
\\  
 $^1$ Inter-University Centre for Astronomy \& Astrophysics, Ganeshkhind, Post Bag 4, Pune 411007, India\\
 $^2$ Center for Particle Cosmology, University of Pennsylvania, 209 S. 33rd St., Philadelphia, PA 19104, USA\\
 $^3$ The Abdus Salam International Center for Theoretical Physics, Strada Costiera, 11, Trieste 34151, Italy      }

\begin{document}
\label{firstpage}
\pagerange{\pageref{firstpage}--\pageref{lastpage}}
\maketitle

\begin{abstract}
In general relativity (GR), the internal dynamics of a self-gravitating system under free-fall in an external gravitational field should not depend on the external field strength.  
Recent work has claimed a statistical detection of an `external field effect' (EFE) using galaxy rotation curve data.  
We show that large uncertainties in rotation curve analyses and inaccuracies in published simulation-based external field estimates compromise the significance of the claimed EFE detection. 
We further show analytically that a qualitatively similar statistical signal is, in fact, expected in a $\Lambda$-cold dark matter ($\Lambda$CDM) universe without any violation of the strong equivalence principle. 
Rather, such a signal arises simply because of the inherent correlations between galaxy clustering strength and intrinsic galaxy properties. 
We explicitly demonstrate the effect in a baryonified mock catalog of a $\Lambda$CDM universe.  
Although the detection of an EFE-like signal is not, by itself, evidence for physics beyond GR, our work shows that the \emph{sign} of the EFE-like correlation between the external field strength and the shape of the radial acceleration relation
can be used to probe new physics:  e.g., in MOND, the predicted sign is opposite to that in our $\Lambda$CDM mocks. 
\end{abstract}


\section{Introduction}
\label{sec:intro}
\noindent
The rotation curves of spiral galaxies, and more recently the velocity dispersion profiles of elliptical galaxies, show that the acceleration \atot\ one infers from the observed motions of their stars or cold gas differs from the acceleration \abary\ which one estimates from their observed baryonic mass distribution, if one assumes the motions are driven by Newtonian gravity.  Nevertheless, the two accelerations define a rather tight correlation \citep{mls16,janz+16,lmsp17,cbsg19,tian+20,efe20}, which is known as the radial acceleration relation (hereafter RAR).  In cold dark matter (CDM) dominated models, Newtonian gravity is an excellent approximation, so both the shape and tightness of the RAR must emerge from the mixing of the baryonic and dark matter components as a galaxy's stars form and its mass is assembled.  In Modified Newtonian Dynamics \citep[MOND,][]{milgrom83-MOND1,bm84}, which assumes there is no dark matter component, the RAR is a consequence of the departure from the Newtonian force law at small accelerations $|\mathbf{a}|\ll a_0$, with $a_0\sim10^{-10}\msq$ being a fundamental acceleration scale postulated in the theory. So, while the exact shape of the RAR depends on precisely how the gravitational force is modified, its tightness is `natural'.  Both approaches are able to describe the observed shape and tightness of the RAR 
(see, e.g., \citealp{dcl16,desmond17,navarro+17,ps21} and references therein).
The observed median RAR is well described by 
$\atot/\abary = {\cal F}(\abary/a_0)$ where
\be
\Cal{F}(x) = \left[\,\frac12 + \sqrt{\frac14+\frac{1}{x^\nu}}\,\right]^{1/\nu}\,,
\label{eq:simpleIF}
\ee
with $a_0=1.2\times10^{-10}\msq$ and $\nu\simeq0.8$-$1$ \citep{cbsg19,efe20}.

In general relativity (GR), the central assumption of the strong equivalence  principle (SEP) means that the internal dynamics of a self-gravitating system under free-fall in an external gravitational field does not depend on the strength of the external field.  However, MOND violates the SEP \citep[see][who presented a Lagrangian formulation of the theory]{bm84}  and consequently predicts an external field effect \citep[EFE,][]{milgrom83-MOND1}.
Consider a self-gravitating object, such as a star in a galaxy or a galaxy in the cosmic web, which experiences an external gravitational field $\aext$ (whose tidal influence is assumed to be negligible).
At sufficiently large distances from the object and in its center-of-mass rest frame, the EFE manifests as an effective dependence of Newton's constant $G_{\rm N}$ on  $\aext$, with the gravitational force experienced  by a test particle in this frame being approximately Newtonian (and not MOND-ian) but with $G_{\rm N}=G_{\rm N}(|\aext|)$ \citep[e.g., equation~32 of][]{bm84}. If $|\aext|\ll a_0$, the modified $G_{\rm N}$ scales like $\sim a_0/|\aext|$.

Generic solutions of the field equations of MOND (and hence the EFE) relevant for galactic rotation curves have also been discussed in the literature (see, e.g., \citealp{milgrom86} for early work and \citealp{fm12} for a recent review). While the 3-dimensional case requires numerical integration, in 1 dimension one can derive analytical solutions \citep[see, e.g., section~6.3 and equation~59 of][]{fm12}. These have been used in the literature as heuristic approximations to search for observational signatures of the EFE \citep{lelli+15,hbzk16,haghi+19,efe20}.

Recent work has claimed a statistical detection \citep[][hereafter, C20;  see also \citealp{efe-erratum} and \citealp{efe21}]{efe20} in a subsample of 148 disk galaxies taken from the SPARC sample \citep{lms16b}.
This detection boils down to noticing 
 \begin{enumerate}
  \item a systematic departure from the RAR at low accelerations if no EFE is assumed (i.e. from equation~\ref{eq:simpleIF}), and 
  \item a correlation of this departure with the external environment. 
 \end{enumerate}
The EFE predicts that the RAR of an individual galaxy will deviate from \eqn{eq:simpleIF} by an amount determined by $\aext$; in the heuristic approaches mentioned  above, this departure is a dip below \eqref{eq:simpleIF} that becomes larger as the strength $|\aext|$ of the external field increases.

The main goal of the present study is to show that a \emph{statistical} EFE-like signal is, in fact, generically expected in CDM models when using realistic galaxy rotation curves. For this purpose, in addition to analytical arguments, we will use a mock galaxy catalog which \citetalias{ps21} showed is able to describe the other aspects of the RAR. We will focus on the low-acceleration regime $\abary\leq10^{-10}\msq$ where the difference between setting $\nu=0.8$ or $\nu=1$ in \eqn{eq:simpleIF} is negligible. The paper is organised as follows. In section~\ref{sec:analytical}, we present a simple but generic analytical calculation which demonstrates the existence of a statistical EFE in CDM models. In section~\ref{sec:numerical}, we describe our mock catalog, and present our numerical results.  Section~\ref{sec:compare} presents a comparison with the literature, highlighting differences in how measurement uncertainties are incorporated into the analysis, and how the external field strength is estimated.  We conclude in section~\ref{sec:conclude}.  

Throughout, $R_{\rm vir}$ refers to the spherical radius around the host halo center-of-mass which encloses a total matter density $200$ times the critical density $\rho_{\rm crit}$ of the  Universe, while  $m_{\rm vir}$ denotes the total mass enclosed inside this radius.

\section{Analytical expectations}
\label{sec:analytical}
To understand what an `external field effect' might look like in CDM models, we calculate the acceleration on the galaxy in question (which resides in a host halo of radius $R_{\rm vir}$) due to the mass external to $R_{\rm vir}$.

Consider an arbitrary external matter distribution with overdensity $\Delta(\mathbf{r})=1+\delta(\mathbf{r})=\rho(\mathbf{r})/\bar\rho$ given by  the multipole expansion
\be
\Delta(\mathbf{r})-1 = \Theta(r-R_{\rm vir})\sum_{\ell=0}^\infty\sum_{m=-\ell}^\ell\Delta_{\ell m}(r)Y^m_\ell(\hat r)\,,
\ee
where $\Theta(x)$ is the Heaviside step function  and $Y^m_\ell(\hat r)$ are spherical harmonics.\footnote{Our convention is formally equivalent to assuming a uniform matter density at $r<R_{\rm vir}$. It is straightforward to replace this with any other internal distribution if needed, ensuring appropriate boundary conditions at $r=R_{\rm vir}$, without affecting the conclusions regarding \aext.} One can derive the following exact relation for \aext\ by solving Poisson's equation $\nabla^2\phi^{\rm (ext)}=4\pi G\bar\rho(\Delta-1)$ for the potential $\phi^{\rm (ext)}$ in spherical polar coordinates \citep[e.g.,][]{binney-tremaine-GalDyn} and evaluating its gradient at the origin:
\begin{align}
&\aext = -\nabla\phi^{\rm (ext)}(\mathbf{r}=0) \notag\\
&= \sqrt{\frac{3}{16\pi}}\,\Omega_{\rm m}H^2\left[\hat z\,\Cal{D}_{10} - \sqrt{2}\left(\hat x\,{\rm Re}(\Cal{D}_{11})-\hat y\,{\rm Im}(\Cal{D}_{11})\right)\right]\,,
\label{eq:aext-exact}
\end{align}
where we defined the integrals
\be
\Cal{D}_{1m}  \equiv \int_{R_{\rm vir}}^\infty\der r\,\Delta_{1m}(r)\,,
\label{eq:D1m-def}
\ee
used the fact that $\Delta_{10}$ is real, $\Delta_{1,-1} =  -\Delta_{11}^\ast$ and made the associations $\hat r=\hat z$, $\hat\theta=\hat x$ and $\hat\phi=\hat y$ at the origin of coordinates. Thus, for an arbitrary inhomogeneous external matter distribution, it is only the $\ell=1$ terms that contribute to an external field at the origin. The remaining terms vanish either due to symmetry (as in the case of the monopole $\ell=0$) or because they scale like positive powers of $r\to0$. This trivially recovers the well-known result that \aext\ at the origin in Newtonian gravity vanishes for a perfectly spherical external mass distribution. It also shows that an axisymmetric dipolar mass field with $\delta(\mathbf{r})=\Delta_{10}(r)Y^0_1(\hat r)\sim\Delta_{10}(r)\cos(\theta)$ will lead to an external field $\aext$ aligned with the dipole axis $\hat  z$.

Dynamically, of course, this still does not explain why internal quantities such as the rotation curve of the galaxy should  depend on \aext\ (apart from a trivial dependence on the chosen boundary at $r=R_{\rm vir}$).
To see what \eqn{eq:aext-exact} implies \emph{statistically}, we first relate the multipole coefficients of the external matter field to its Fourier transform. Using the orthogonality of the spherical harmonics and the multipole expansion of the exponential $\e{i\mathbf{k}\cdot\mathbf{r}}=4\pi\sum_{\ell=0}^\infty i^\ell j_\ell(kr)\sum_{m=-\ell}^\ell Y^{m\ast}_\ell(\hat k)Y^m_\ell(\hat r)$, where $j_\ell$ are spherical Bessel functions, leads to the relation $\Delta_{\ell m}(r)=4\pi i^\ell\int\der^3k/(2\pi)^3\,\delta_{\mathbf{k}}\,j_\ell(kr)\,Y^{m\ast}_\ell(\hat k)$ in terms of the Fourier transform $\delta_{\mathbf{k}}=\int\der^3r\,\e{-i\mathbf{k}\cdot\mathbf{r}}\delta(\mathbf{r})$. This in turn gives
\be
\Cal{D}_{1m} = 4\pi  i\int\frac{\der^3k}{(2\pi)^3}\,Y^{m\ast}_1(\hat k)\,k^{-1}\,j_0(kR_{\rm vir})\,\delta_{\mathbf{k}}\,,
\ee
where we used $\int_A^\infty\der x\,j_1(x)=j_0(A)$, so that 
\be
\avg{\Cal{D}_{1m}\Cal{D}_{1m}^\ast} = \frac2\pi\int_0^\infty\der k\,j_0(kR_{\rm vir})^2\,P_{\rm mm|g}(k)\,,
\ee
independent of $m$, where we used $\int\der\Omega_k\,|Y^m_\ell(\hat k)|^2=1$ and where $P_{\rm mm|g}(k)$ is the power spectrum of the mass external to the galaxy (i.e., conditioned on there being a galaxy at the center). This finally leads to the expectation value,
\be
\avg{\aext\cdot\aext} = \left(\frac{3\Omega_{\rm m}H^2}{2}\right)^2\int_0^\infty\frac{\der k}{2\pi^2}\,j_0(kR_{\rm vir})^2\,P_{\rm mm|g}(k)\,.
\label{eq:<aext^2>-exact}
\ee
Although formally a 2-point quantity, $P_{\rm mm|g}(k)$ is essentially a galaxy-mass-mass bispectrum, and we finally see why rotation curves might be expected to correlate with \aext. This is simply because rotation curves depend on galaxy properties such as host halo mass, \emph{as does} the large-scale clustering implied by $P_{\rm mm|g}(k)$, e.g., through the galaxy's linear bias $b_1$ which we discuss below \citep[see][for a review]{djs18}. CDM models therefore implicitly contain a \emph{statistical} external field effect.

To see how this manifests in the RAR, we turn to a numerical study using a mock catalog of galaxy rotation curves in the next section.

\section{Numerical results}
\label{sec:numerical}

\subsection{Mock catalog}
\label{subsec:mock}
Our mock catalog is the same as used by \citetalias{ps21} and is based on the algorithm described by \citet[][hereafter, PCS21]{pcs21}. We briefly describe the key features of the mock relevant to our analysis here, and refer the reader to \citetalias{ps21} and \citetalias{pcs21} for further details of the mock algorithm and underlying $N$-body simulation.

The catalog represents a luminosity-complete sample of galaxies with $r$-band absolute magnitude $M_r\leq-19$ in a $(300\Mpch)^3$ comoving volume at $z=0$. The mock contains both central and satellite galaxies, populated in dark matter  haloes identified in an $N$-body simulation having $1024^3$ particles with a flat $\Lambda$CDM WMAP7 cosmology (\citealp{komatsu+2011}; $\Omega_{\rm m}=0.276$, $h=0.7$). To achieve the luminosity completeness threshold of $M_r\leq-19$, haloes containing $\geq40$ particles are considered. Since the halo concentration cannot be reliably measured for haloes with fewer than about $\sim400$ particles, we use the method presented by \citet{rps21} to assign concentrations $c_{\rm vir}$ \citep[assuming][NFW profiles]{nfw96} conditioned on the mass and local tidal environment of individual haloes.\footnote{The calibration of \citet{rps21} assumes concentrations $c_{\rm 200b}\equiv R_{\rm 200b}/r_{\rm s}$, where $R_{\rm 200b}$ is the halo-centric radius which encloses a density 200 times the background value and $r_{\rm s}$ is the halo's NFW scale radius. We convert these to $c_{\rm vir}\equiv R_{\rm vir}/r_{\rm s}$ using the analytical prescription of \citet{hk03}.} The galaxies were populated using a halo occupation distribution (HOD) model and \Hi-optical scaling relations calibrated by \citet{pcp18} and \citet{ppp19} using luminosity- and colour-dependent clustering measurements from the Sloan Digital Sky Survey \citep[SDSS,][]{york+00,zehavi+11} and \Hi-dependent clustering measurements from the Arecibo Legacy Fast ALFA survey \citep[ALFALFA,][]{giovanelli+05,guo+17}. Each galaxy in the mock is assigned absolute magnitudes in SDSS $u$, $g$ and $r$ bands (with a threshold $M_r\leq-19$ imposed by the SDSS clustering measurements) and a stellar mass $m_\ast$ using a colour-dependent mass-to-light ratio. The \Hi-optical scaling relation additionally leads to a fraction $\sim60\%$ of galaxies to be  assigned an \Hi\ mass $m_{\Hi}$. \citetalias{pcs21} presented extensive tests of this algorithm. 
We focus in this work on the population of mock central galaxies containing massive \Hi\ disks, with $m_{\Hi}\geq10^{9.7}\Mhsq$; the resulting $\sim50,000$ such objects in our catalog form a volume-complete sample of \Hi-selected galaxies.

The host haloes of these centrals are `baryonified' by the \citetalias{pcs21} algorithm using a modified version of the prescription of \citet{st15}. In addition to optical luminosity, colour, stellar mass and (where available) \Hi\ mass,  each central galaxy system is further assigned spatial distributions of stars, cold gas, hot ionised gas, and gas `expelled' due to feedback processes. The shapes of these distributions  are observationally constrained; of relevance below are the distributions of stars and cold gas, which are respectively modelled as a Gaussian sphere\footnote{We have verified that modelling the stellar component as the combination of a central bulge and a thin disk, with a fixed bulge-to-total mass ratio $B/T<1$, does not qualitatively affect any of our conclusions. We have also checked that including a dependence of $B/T$ on properties such as $m_{\rm vir}$ only affects the high-$\abary$ end of the RAR, which is irrelevant for the present work. We will report the results of including the effects of a realistic $B/T$ distribution on the RAR in future work.} with half-light radius $R_{\rm hl}\propto R_{\rm vir}$ and an axisymmetric thin exponential disk with scale radius $h_{\Hi}\propto m_{\Hi}^{0.5}$ (see \citetalias{pcs21} and \citetalias{ps21} for details and original references).

Finally, as discussed in detail by  \citetalias{ps21}, an important aspect of this `baryonification' scheme as regards RAR studies is that the dark matter component of each baryonified halo is allowed to respond to the presence of its baryonic distribution by modelling a quasi-adiabatic relaxation process \citep[][see section~3 of \citetalias{ps21}]{teyssier+11,st15}. This is done in a parametrised fashion using a `relaxation parameter' $q_{\rm rdm}$ to  control the amount of quasi-adiabatic relaxation, such that $q_{\rm rdm}=1$ ($q_{\rm rdm}=0$) corresponds to perfect angular momentum conservation (no baryonic backreaction). The value $q_{\rm rdm}=0.68$ provides a good description of the effect seen in cluster-sized haloes in hydrodynamical CDM zoom simulations \citep{teyssier+11,st15}, but is subject to some theoretical uncertainty as discussed by \citetalias{ps21}, settling which requires a detailed study of hydrodynamical simulations of galaxy formation over a large mass range. 
In the following, we set $q_{\rm rdm}=0.33$ when generating baryonified rotation curves; \citetalias{ps21} showed that this improves the agreement at high accelerations $\abary\geq10^{-10}\msq$ between the median RAR of our mock and \eqn{eq:simpleIF} with $\nu=1$ (which \citealp{efe20} argue provides a good description of the SPARC sample at similar \abary). At the low accelerations of our interest, however, our results are insensitive to $q_{\rm rdm}$ and are unchanged if we use the default value $q_{\rm rdm}=0.68$ used by \citetalias{ps21}.

It is also interesting to ask which other aspects of the modelling choices outlined above might affect our subsequent results. While the HOD and associated scaling relations underlying our mock catalog are tightly constrained by the SDSS and ALFALFA data \citep{pcp18,ppp19}, the results of \citetalias{ps21} show that the modelling of `expelled' gas, or the circum-galactic medium (CGM), can have interesting effects on the RAR shape in the outer halo. Since the distribution of the CGM is observationally ill-constrained, it will be very interesting to understand the sensitivity of our results below to CGM modelling choices. We leave this exercise to future work.

\subsection{RAR and the environment in a CDM mock catalog}
\label{subsec:mockEFE}
\begin{figure}
\centering
\includegraphics[width=0.48\textwidth,trim=5 10 0 5,clip]{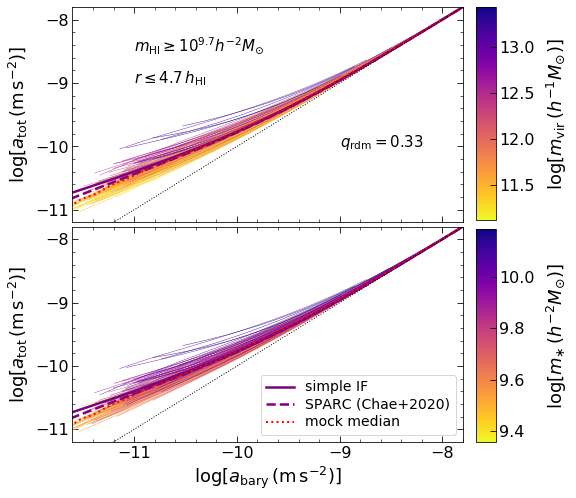}
\caption{{\bf RAR and halo/stellar mass}. Radial acceleration relation (RAR) for $150$ massive spiral galaxies selected randomly from a volume-complete mock catalog with $M_r\leq-19$ and $m_{\Hi}\geq10^{9.7}\Mhsq$. Thin curves show the RARs of individual galaxies coloured by host halo mass $m_{\rm vir}$ \emph{(top panel)} and stellar mass $m_\ast$ \emph{(bottom panel)}.  
Each rotation curve is truncated at $r\leq4.7h_{\Hi}$ to mimic an observational column density threshold of $N_{\Hi}\geq10^{19.5}\,{\rm cm}^{-2}$. 
In each panel, the solid purple curve shows \eqn{eq:simpleIF} setting $\nu=1$, the dashed purple curve shows equation~(6) of \citetalias{efe20} setting their parameter $e=0.032$, the dotted red curve shows the median RAR of the full volume-complete sample (containing $\sim50,000$ galaxies) and the thin dotted black line shows the 1:1 relation. 
See text for a discussion. 
}
\label{fig:efe-downturn}
\end{figure}

The observable that \citetalias{efe20} attribute to an EFE is a downward deviation of the RAR of individual galaxies or the ensemble from \eqn{eq:simpleIF}.
\citetalias{ps21} showed that, in CDM models, the objects which dominate the downturn in the RAR tend to have smaller masses. This is highlighted in Fig.~\ref{fig:efe-downturn} which shows the individual RARs of $150$ galaxies randomly chosen from our full sample. Each curve in the \emph{upper panel} is coloured by the total mass $m_{\rm vir}$ of the host halo, while the \emph{lower panel} shows the same curves coloured by the stellar mass $m_\ast$ of each galaxy. We clearly see the low-$m_{\rm vir}$ objects in the \emph{upper panel} falling below \eqn{eq:simpleIF} (shown as the solid purple curve in each panel), while this correlation visibly weakens when using $m_\ast$ in place of $m_{\rm vir}$ in the \emph{lower panel}. This weakening is not surprising, considering the substantial scatter of the stellar mass-halo mass relation \citepalias[see, e.g., fig.~12 of][]{pcs21}. The rotation curves used for evaluating \atot\ and \abary\ are sampled on 20 logarithmically spaced points in the range $(10^{-3},1)\times R_{\rm vir}$ for each  galaxy. We have displayed each curve in the range $r\leq4.7h_{\Hi}$, where $h_{\Hi}$ is the scale length of the thin exponential \Hi\ disk assigned to each galaxy, with surface mass density of \Hi\ gas $\Sigma_{\Hi}(r)\propto\e{-r/h_{\Hi}}$ in the disk plane. Since our model assumes $h_{\Hi}\propto m_{\Hi}^{0.5}$, this cut on $r/h_{\Hi}$ corresponds to a column density threshold of $N_{\Hi}\geq10^{19.5}\,{\rm cm}^{-2}$, which follows from writing $N_{\Hi}(r)\simeq\Sigma_{\Hi}(r)/m_{\rm p}$ ($m_{\rm p}$ being the proton mass) and represents a typical $5\sigma$ limiting threshold for spatially resolved 21 cm spectroscopy \citep[e.g.,][]{bc04,bfos06,ccdf06,boomsma+08}.

The dashed purple curve in each panel shows the RAR derived from equation~(6) of \citetalias{efe20}, setting their parameter $e=0.032$,  which they showed describes the median  RAR of the SPARC sample at $\abary\lesssim10^{-11}\msq$ somewhat better than does \eqn{eq:simpleIF}. For comparison, the dotted red curve in each panel shows the  median RAR of our entire volume-complete mock sample. Like the dashed purple curve, the median RAR of our mock also dips below \eqn{eq:simpleIF} at low \abary. The RARs of the individual galaxies show that this is driven by the low-mass host halos. The individual RARs also depend on halo concentration: we discuss this in more detail in the next subsections.  For reference, the median along with $16^{\rm th}$ and $84^{\rm th}$ percentiles of $\log[m_{\rm vir}(\Mh)]$ for this sample are $11.88_{-0.44}^{+0.66}$. We have checked that our results are robust against varying the cut on $r/h_{\Hi}$ between $\sim3.5$-$6$, corresponding to column density thresholds of $\sim10^{19}$-$10^{20}\,{\rm cm}^{-2}$.

\subsubsection{RAR and large-scale halo bias}
Since halo mass and concentration in the CDM paradigm correlate with large-scale environment, the discussion above shows that we would also expect the galaxies dipping below \eqn{eq:simpleIF} in Fig.~\ref{fig:efe-downturn} to have smaller values of linear bias $b_1$.
We test this expectation as follows.  
As a proxy for the EFE observable, for each object in the mock catalog, we first estimate
\be 
\varepsilon \equiv \avg{\left[\frac{\atot}{a_{\rm tot,med}(\abary)} - 1 \right] }\,,
\label{eq:eps-def}
\ee
where $a_{\rm tot,med}(\abary)$ is the median \atot\ measured in narrow bins of \abary\ using all galaxies in our sample, and interpolated  to the value of \abary\ for each galaxy. The angular brackets indicate, for each galaxy, the median over data points for which $10^{-12}\leq\abary/(\msq)\leq10^{-10}$ and $r\leq4.7\,h_{\Hi}$ (see above). Using the mean or minimum instead of the median in defining the angular brackets leads to very similar results, as does varying the cut on $r/h_{\Hi}$ between $\sim3.5$-$6$ (see above). 
Since $\varepsilon$ is averaged over a wide range of \abary, it is a measure of the overall offset of a galaxy's RAR from the sample median, and does not distinguish between differences in shape or amplitude of the RAR. In contrast, the parameter $e$ used by \citetalias{efe20} and which we discuss later, is intended to quantify differences in RAR shape, although in practice the best-fit $e$ for any galaxy may depend on whether the acceleration scale $a_0$ is also treated as a free parameter.

\begin{figure}
\centering
\includegraphics[width=0.45\textwidth]{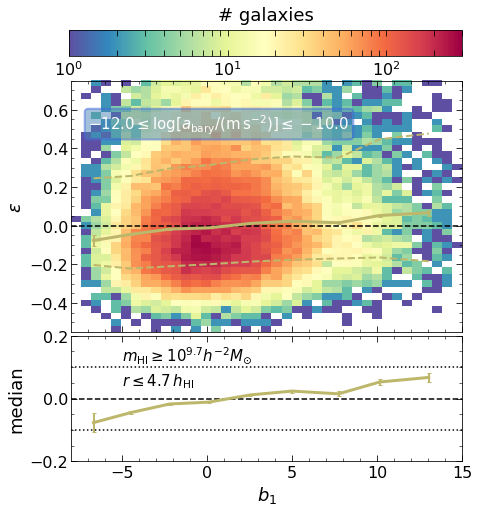}
\caption{{\bf Statistical EFE in CDM.} \emph{(Top panel):} Joint distribution of $\varepsilon$ defined in \eqn{eq:eps-def}, which measures a departure from the median RAR, and galaxy linear bias $b_1$ in the same mock catalog containing $\sim50,000$ galaxies used for Fig.~\ref{fig:efe-downturn}.
Dashed black horizontal line shows the median value of $\varepsilon$ (essentially zero) in the mock catalog. Solid yellow line shows the median $\varepsilon$ in bins of $b_1$, with error bars calculated using $150$ bootstrap samples. Dashed yellow lines show the $16^{\rm th}$ and $84^{\rm th}$ percentiles of $\varepsilon$ in the same $b_1$ bins.
\emph{(Bottom panel):} Zoom-in view of the solid yellow line. Dotted horizontal lines show $\pm10\%$ variations around zero (shown by the dashed horizontal line).
A weak but significant positive trend is apparent in the median $\varepsilon$ as a function of $b_1$. 
}
\label{fig:efe-b1}
\end{figure}

In addition, for each object we estimate the linear bias $b_1$, following \cite{phs18}, as a proxy for the large-scale environment. 
The \emph{top panel} in Fig.~\ref{fig:efe-b1} shows $\varepsilon$ vs $b_1$ for our sample (all these objects have at least one value of \abary\ between $10^{-12}-10^{-10}\msq$ such that $r\leq4.7\,h_{\Hi}$). 
The solid yellow line shows the median $\varepsilon$ for narrow bins in $b_1$, and dashed yellow lines show the region which encloses 68\% of the objects.  There is a clear trend with $b_1$, which the \emph{bottom panel} zooms in on.  Evidently, the median $\varepsilon$ in the most overdense environments is weakly but significantly higher than that in the most underdense ones, with a $\sim15\%$ overall change from $b_1\sim-7$ to $b_1\sim12$. Mean density environments have median $\varepsilon\simeq0$. We find a Spearman correlation coefficient of $+0.08$  between $\varepsilon$ and $b_1$, with negligible $p$-value, consistent with the median trend. This qualitative trend is consistent with the expectations from the RAR analysis mentioned above: galaxies which dip below \eqn{eq:simpleIF} have preferentially smaller values of $b_1$.

We have also explicitly checked that the trend between $\varepsilon$ and $b_1$ disappears when evaluated at fixed  host mass $m_{\rm vir}$ and concentration $c_{\rm vir}$. We do this by rank-ordering $b_1$ in joint percentiles of $m_{\rm vir}$ and $c_{\rm vir}$ and correlating the resulting ranks of $b_1$ with $\varepsilon$. The resulting median $\varepsilon$ is consistent with zero at fixed $b_1$ rank, with errors similar to or smaller than those displayed in Fig.~\ref{fig:efe-b1}, and the Spearman correlation coefficient between $\varepsilon$ and the $b_1$ rank is $+0.006$ with a $p$-value of $0.17$, indicating no significant correlation. This emphasizes that $\varepsilon$ and $b_1$ are only correlated because each of them separately correlates with $m_{\rm vir}$ and $c_{\rm vir}$. Interestingly, we also find that fixing $m_{\rm vir}$ alone decreases, but does not completely erase,  the $\varepsilon\leftrightarrow b_1$ correlation (Spearman correlation $+0.05$ with negligible $p$-value), showing that halo assembly bias effects can leave (weak) imprints in RAR-environment correlations. We return to this point below.

\begin{figure}
\centering
\includegraphics[width=0.48\textwidth]{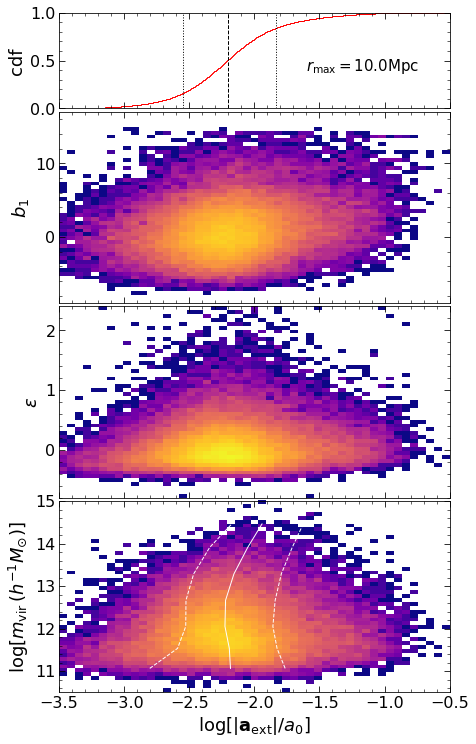}
\caption{{\bf Distribution of $|\aext|$ for mock galaxies.} \emph{Upper-most panel} shows the cumulative distribution of $\log[|\aext|/a_0]$ estimated using $r_{\rm max}=10\,{\rm Mpc}$ as described in the text for the same mock galaxies used in Fig.~\ref{fig:efe-b1}. Vertical dashed and dotted lines indicate the median and central $68\%$ region, respectively, of the distribution. Subsequent panels show, from \emph{top} to \emph{bottom}, the joint distribution of $\log[|\aext|/a_0]$ with linear bias $b_1$, the RAR residual $\varepsilon$ from \eqn{eq:eps-def} and host mass $\log[m_{\rm vir}]$, with the colour indicating binned galaxy counts on a logarithmic scale (decreasing from yellow to purple). White solid line in the bottom panel shows the median $\log[|\aext|/a_0]$ in bins of $\log[m_{\rm vir}]$, with dashed white lines showing the corresponding central $68\%$ region. 
}
\label{fig:aextstats}
\end{figure}

\begin{figure*}
\centering
\includegraphics[width=0.98\textwidth,trim=10 10 5 5,clip]{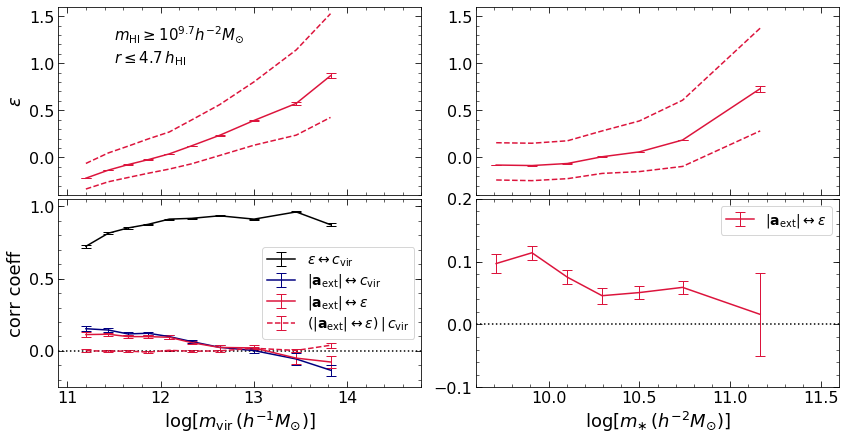}
\caption{{\bf Correlation between $\varepsilon$ and $|\aext|$ for mock galaxies.} \emph{(Top left panel:)} Median (solid red) and central $68\%$ region (dashed red) of the distribution of $\varepsilon$ in bins of $\log[m_{\rm vir}]$ for the same mock galaxies used in Fig.~\ref{fig:efe-b1}. Both the median and the scatter in $\varepsilon$ monotonically increase with $m_{\rm vir}$. 
\emph{(Bottom left panel:)} Solid lines show Spearman rank correlation coefficients between $\varepsilon\leftrightarrow c_{\rm vir}$ (black), $|\aext|\leftrightarrow c_{\rm vir}$ (blue) and $|\aext|\leftrightarrow \varepsilon$ (red) in bins of $m_{\rm vir}$. The latter two curves show a clear inversion of sign near $m_{\rm vir}\simeq10^{13}\Mh$, which the text argues is a version of halo assembly bias.
Dashed red line shows the conditional correlation coefficient $\gamma_{(|\aext|\leftrightarrow \varepsilon)|c_{\rm vir}}$ as defined in \eqn{eq:cond-cc}. This is nearly zero across the entire range of $m_{\rm vir}$, an indication that the $|\aext|\leftrightarrow \varepsilon$ correlation is driven by halo concentration $c_{\rm vir}$. 
\emph{(Top right panel:)} Median (solid red) and central $68\%$ region (dashed red) of the distribution of $\varepsilon$ in bins of log-stellar mass $\log[m_{\ast}]$ for the same mock galaxies. The trend with $m_\ast$ is much shallower than that with $m_{\rm vir}$, a consequence of the scatter in the $m_\ast$-$m_{\rm vir}$ relation (see text).
\emph{(Bottom right panel:)} $|\aext|\leftrightarrow \varepsilon$ correlation in bins of $\log[m_\ast]$. This remains positive for nearly all $m_\ast$ explored in the mock.
Error bars in all panels were computed as the scatter across $150$ bootstrap realisations.}
\label{fig:aext-epsilon}
\end{figure*}

\subsubsection{RAR and the external acceleration field}
In the context of \eqn{eq:<aext^2>-exact}, we  expect that $\varepsilon$ must also correlate with $|\aext|$. To test this, we measured $\aext$ at each host halo location in our simulation box as the contribution of all mass in the radial shell $R_{\rm vir}\leq r\leq r_{\rm max}$. For ease of comparison with \citet[][hereafter, D18]{dflj18} who performed a similar study with a different technique, we set $r_{\rm max}=10\,{\rm Mpc}$ for our default analysis and comment on the scale dependence of our results later. 
In practice, we calculated \aext\ by summing over the vector accelerations induced by all dark matter particles in the radial shell centered on each halo, at the halo center:
\begin{align}
\aext &= \sum_{p\in{\rm shell}} \frac{Gm_{\rm part}\mathbf{x}_p}{x_p^3} = \frac{3\Omega_{\rm m}H^2L}{8\pi N_{\rm part}}\sum_{p\in{\rm shell}}\frac{\mathbf{r}_p}{r_p^3}\,,
\label{eq:aext-sim}
\end{align}
where $\mathbf{r}_p = \mathbf{x}_p/L$ is the halo-centric position vector of the $p^{\rm th}$ particle, normalised by the box size $L$, and $N_{\rm part}$ is the total number of particles in the simulation. To speed up the calculation, we first downsampled the particle distribution to $256^3$ particles (from the native sampling of $1024^3$ particles), replacing $N_{\rm part}\to256^3$ in \eqn{eq:aext-sim}. We have checked using halo-based samples that our results are converged with respect to downsampling level.\footnote{We have also checked, using a $(600\Mpch)^3$ simulation with $1024^3$ particles, that our results for halos with $m_{\rm vir}\gtrsim10^{12}\Mh$ (including the correlations of \aext\ with halo mass and concentration discussed later) are converged with respect to box volume.}
Unlike the discrete halo counting employed by \citetalias{dflj18}, which necessarily requires making assumptions regarding the mass contributed by unresolved halos in the simulation box, our method correctly accounts for \emph{all} mass in the desired radial shell; we return to this point in section~\ref{sec:compare}.

Fig.~\ref{fig:aextstats} shows the cumulative distribution of $\log[|\aext|/a_0]$ with $a_0=1.2\times10^{-10}\msq$ in the \emph{upper-most panel}, followed by the joint distribution of $\log[|\aext|/a_0]$ with $b_1$, $\varepsilon$ and $\log[m_{\rm vir}]$ (2-d histograms from top to bottom). The median $|\aext|/a_0$ along with the central $68\%$ range is $(6.2^{+8.4}_{-3.4})\times10^{-3}$ (vertical lines in the \emph{upper-most panel}). We see a relatively tight correlation between $|\aext|$ and $b_1$ (Spearman correlation coefficient $\simeq+0.18$ with negligible $p$-value), consistent with expectations from \eqn{eq:<aext^2>-exact}. The corresponding correlation between $|\aext|$ and $\varepsilon$ (Spearman coefficient $\simeq+0.08$ with negligible $p$-value) is similar to the one between $b_1$ and $\varepsilon$ discussed above. The \emph{lower-most panel} shows that $|\aext|$ and host mass $m_{\rm vir}$ define a tight correlation at large $m_{\rm vir}$, which is then inherited by $\varepsilon$ through its $m_{\rm vir}$ dependence.

Fig.~\ref{fig:aext-epsilon} explores the $|\aext|\leftrightarrow\varepsilon$ correlation in more detail. Since the RAR residual $\varepsilon$ correlates with the environment through its dependence on halo mass $m_{\rm vir}$ and concentration $c_{\rm vir}$, it is natural to ask how its correlation with $|\aext|$ is affected by these variables. For reference, the \emph{top left panel} of the Figure shows the median and the region containing the central 68 percent of $\varepsilon$ values in bins of $m_{\rm vir}$, while the solid black curve in the \emph{bottom left panel} shows the Spearman rank correlation coefficient $\gamma_{\varepsilon c_{\rm vir}}$ between $\varepsilon$ and $c_{\rm vir}$; we see a clear monotonic increase in both the median and scatter of $\varepsilon$ with $m_{\rm vir}$ and a strong positive $\varepsilon\leftrightarrow c_{\rm vir}$ correlation at fixed $m_{\rm vir}$, as expected from the discussion in \citetalias{ps21} (see their fig.~6). 
The solid red line in the \emph{bottom left panel} of Fig.~\ref{fig:aext-epsilon} shows the Spearman  coefficient $\gamma_{|\aext|\varepsilon}$ between $|\aext|$ and $\varepsilon$ at fixed $m_{\rm vir}$. We see a weak but significant correlation which is positive at low $m_{\rm vir}$ and changes sign at $m_{\rm vir}\gtrsim10^{13}\Mh$.\footnote{We have checked that the values of $\gamma_{|\aext|\varepsilon}$ vary by only a few percent at any $m_{\rm vir}$ when the cut on $r/h_{\Hi}$ is varied between $\sim3.5$-$6$ (see section~\ref{subsec:mockEFE}).} This change of sign is reminiscent of the well-known $b_1\leftrightarrow c_{\rm vir}$ assembly bias correlation at fixed $m_{\rm vir}$, which similarly changes sign from positive to negative at $m_{\rm vir}\gtrsim10^{13}\Mh$ \citep{wechsler+06,fw10}. Indeed, the solid blue line shows that the Spearman coefficient $\gamma_{|\aext|c_{\rm vir}}$ between $|\aext|$ and $c_{\rm vir}$ also shows the same behaviour with $m_{\rm vir}$. Since $\varepsilon$ in our $\Lambda$CDM mock does not directly depend on the external environment, it is worth asking whether the \emph{entire} $|\aext|\leftrightarrow\varepsilon$ correlation can be explained by the separate correlations $|\aext|\leftrightarrow c_{\rm vir}$ and $\varepsilon\leftrightarrow c_{\rm vir}$, at fixed $m_{\rm vir}$ (just like the $b_1\leftrightarrow\varepsilon$ correlation discussed earlier). To test this, we follow \citet{rphs19} and calculate the conditional correlation coefficient $\gamma_{(|\aext|\leftrightarrow\varepsilon)|c_{\rm vir}}$ defined as
\be
\gamma_{(|\aext|\leftrightarrow\varepsilon)|c_{\rm vir}} = \gamma_{|\aext|\varepsilon} - \gamma_{|\aext|c_{\rm vir}}\,\gamma_{\varepsilon c_{\rm vir}}\,,
\label{eq:cond-cc}
\ee
which should vanish if $|\aext|$ and $\varepsilon$ are only correlated because of their individual correlations with $c_{\rm vir}$. The dashed red line in the \emph{bottom left panel} shows that this is indeed the case. This result is striking in its similarity to that for the $b_1\leftrightarrow\varepsilon$ correlation, and shows the potential of the $|\aext|\leftrightarrow\varepsilon$ correlation in hunting for galaxy assembly bias.\footnote{This result also opens the door to investigating the \emph{origin} of the $|\aext|\leftrightarrow c_{\rm vir}$ correlation. The results of \citet{rphs19}, and the very definition of \aext\ as a derivative of the gravitational potential, suggest that this correlation might ultimately be explained by the tidal environment of the galaxy's host halo. We will pursue this question in future work.}

\begin{figure}
\centering
\includegraphics[width=0.48\textwidth,trim=10 8 8 5,clip]{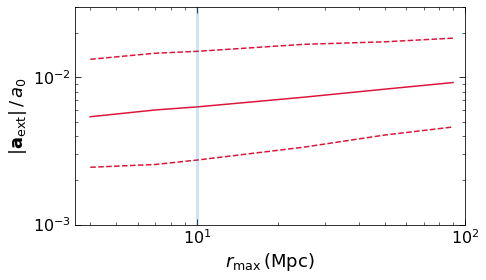}
\caption{{\bf Scale dependence of $|\aext|$} for the mock galaxies used in Fig.~\ref{fig:aextstats}. Solid curve shows the median $|\aext|/a_0$ as a function of $r_{\rm max}$, while the dashed curves show the corresponding central $68\%$ region of the distribution. Vertical line indicates $r_{\rm max}=10\,{\rm Mpc}$, the default value used in the text. 
}
\label{fig:aext-scaledep}
\end{figure}

The sign of the $|\aext|\leftrightarrow\varepsilon$ correlation is thus predicted to depend on $m_{\rm vir}$. In practice, however, one is likely to assess this correlation as a function of quantities such as stellar mass $m_\ast$, which is easier to estimate than halo mass $m_{\rm vir}$. The \emph{bottom right} panel of Fig.~\ref{fig:aext-epsilon} shows that, at fixed $m_\ast$, $\gamma_{|\aext|\leftrightarrow\varepsilon}$ is in fact positive over essentially the entire range of $m_\ast$ probed by our mock catalog. 
Finally, the \emph{top right panel} shows that the distribution of $\varepsilon$ is a much weaker function of $m_\ast$ than it is of $m_{\rm vir}$. This is easy to understand in terms of the shape and scatter of the $m_\ast$-$m_{\rm vir}$ relation (e.g., fig.~12 of \citetalias{pcs21}). At any fixed $m_\ast\lesssim10^{10.5}\Mhsq$, the $\varepsilon$ distribution in our mock catalog is averaged over a similar range of $m_{\rm vir}\sim10^{11.5}$-$10^{12.5}\Mh$, while at $m_\ast\gtrsim10^{10.5}\Mhsq$ it is progressively averaged over larger $m_{\rm vir}$, leading to the steepening seen in the plot.

Overall, then, our $\Lambda$CDM mock catalog predicts that the distribution of $\varepsilon$ is a strong function of $m_{\rm vir}$ but a much weaker function of stellar mass $m_\ast$ and, more interestingly for EFE analyses, that there is a \emph{positive} correlation between $\varepsilon$ and $|\aext|/a_0$ for all systems but those with the highest $m_{\rm vir}$, such that galaxies dipping below \eqn{eq:simpleIF} tend to have \emph{smaller} $|\aext|$. 
This trend between $|\aext|$ and $\varepsilon$ is also different from the MOND prediction where, in the language of \citetalias{efe20}, the variable $e\sim-\varepsilon$ should correlate positively with $e_{\rm env}=|\aext|/a_0$, such that galaxies dipping below \eqn{eq:simpleIF} should have \emph{larger} $|\aext|$. We investigate this issue in the next section.

For completeness, Fig.~\ref{fig:aext-scaledep} shows the dependence of the distribution of $|\aext|$ on the scale $r_{\rm max}$, for the same mock galaxies used in Fig.~\ref{fig:aextstats}. We see a slow rise that extends beyond $r_{\rm max}\gtrsim90\,{\rm Mpc}$. We have also found that the correlation between $\varepsilon$ and $|\aext|$, as measured by the Spearman coefficient, remains approximately constant with $r_{\rm max}$.

\section{Comparison with previous results}
\label{sec:compare}
In this section, we revisit the results of \citetalias{efe20} who parametrised the departure of SPARC galaxy RARs from \eqn{eq:simpleIF} by a dimensionless variable $e$ (determined after fixing the overall scale $a_0$ to the same value for all objects), with positive values indicating downward deviations.  We first recalculate the median $e$ value reported by \citetalias{efe20}.  We then ask whether the individual $e$ values correlate with their estimates of external environment or with host halo masses. 

We obtained $e$ and $e_{\rm env}\equiv|\aext|/a_0$, with $a_0=1.2\times10^{-10}\msq$ from the (corrected) Table~2 of \citet{efe-erratum}.  In what follows, we restrict attention to the 148 SPARC galaxies (of a total of $175$) used by \citetalias{efe20}. In their final analysis, \citetalias{efe20} selected a further subset of $113$ galaxies by demanding that their baryonic accelerations $x_0\sim\log[\abary (\msq)]$ occupy the low acceleration regime, with each galaxy's median $x_0$ required to satisfy $\avg{x_0}\leq-10.3$. This was motivated by the MOND expectation that departures of the RAR from \eqn{eq:simpleIF} should occur only at low accelerations. From the CDM viewpoint, however, there is no reason to exclude data in this manner, so we will show results for the full sample as well as the low-acceleration subset. 

However, before we discuss their work we think it is useful to highlight the fact that there are two separate issues:  
1) What is the $e$-$|\aext|$ correlation in $\Lambda$CDM simulations where \atot\ and \abary\ profiles are known for each object?  E.g., one could, and we believe \citetalias{efe20} should, have done the following:  
Fit the MOND+EFE functional form to each RAR measured in a $\Lambda$CDM simulation to determine an $e$ for each object; measure \aext\ in the same $\Lambda$CDM simulation; so determine the correlation between $e$ and $|\aext|$.  This addresses the question of whether it is correct to assume, as \citetalias{efe20} did, that there is no $e$-$|\aext|$ in $\Lambda$CDM.  Our work with $\varepsilon$ in the previous section strongly suggests otherwise.  

Unfortunately, in real data, we do not know \abary\ or \aext, so we must estimate them.  Therefore, a related but separate question is: 
2) How does one estimate the $e$ - $|\aext|$ correlation in data, in the $\Lambda$CDM context, where one must estimate both \abary\ and \aext?  However this is done, the {\em same} estimation procedure that is used for the data should also be used in the $\Lambda$CDM simulations (where the true correlation is known), since this allows one to check if the procedure produces unbiased estimates of the $e$ - $|\aext|$ correlation.  Although this was not done by \citetalias{efe20}, in what follows, we will revisit their results with both points (1) and (2) in mind.  

\subsection{Median value of $e$}
The simple unweighted median of $e$ values for the $113$ low-acceleration galaxies is $0.052\pm0.017$ (with the error estimated from 100 bootstrap samples).  This agrees with the value and bootstrap error reported by \citetalias{efe20}. Similarly, the unweighted median of their corrected $e_{\rm env}$ values is $0.033\pm0.001$, in agreement with their reported median and error.  However, these estimates do not account for measurement errors.   
Using inverse-variance weighting\footnote{We symmetrised the errors in $e$ by defining $\sigma_e = 0.5\times(\sigma_++\sigma_-)$, where $\sigma_\pm$ are the upper and lower errors reported in the third column of Table~2 of \citet{efe-erratum}. We similarly symmetrised errors in $e_{\rm env}$. The errors in $\log[m_{\rm vir}]$ reported by \citet{llms20} are already symmetric.} 
to do so gives weighted medians 
$$e = 0.008\pm0.017 \qquad {\rm and}\qquad e_{\rm env} = 0.026\pm0.002$$ 
for the \emph{same} objects.  These are important changes, since \citetalias{efe20} used their significantly non-zero value of the median $e$ and its statistical consistency with the median $e_{\rm env}$ to claim a statistical detection of the EFE for the low-acceleration sample. Our weighted median calculations, on the other hand, suggest that $e$ is actually consistent with zero while $e_{\rm env}$ is not (although the large error on $e$ means that it is still consistent with $e_{\rm env}$). 

A similar analysis of the full set of 148 galaxies yields weighted (unweighted) median $e=0.008\pm0.013$ ($0.039\pm0.014$), and $e_{\rm env} = 0.026\pm0.001$ ($0.033\pm0.001$).  I.e., as for the low-acceleration subset, the median $e$ is consistent with zero.  The heterogeneity of the SPARC sample, however, means that the errors quoted on individual $e$ values can be substantially over- or under-estimated, making it essential to explore multiple avenues of statistical analysis (K.-H. Chae, private communication).
While the downturn of the average RAR is visually apparent in fig.~3 of \citetalias{efe20}, and so the average $e$ is likely to be non-zero, because the individual values of $e$ are highly uncertain, quantifying it robustly is complicated. 

\subsection{Strength of and correlations with $\aext$}
In addition to the uncertainty on the inferred average value of $e$, an important difference between our results and those of \citetalias{efe20} is that their values of $e_{\rm env}$ are a factor $\sim4$ larger than those predicted by our $\Lambda$CDM mock (Fig.~\ref{fig:aextstats}). Their $\aext$ estimates were derived from a $\Lambda$CDM-based density reconstruction in the observed volume around the SPARC sample following \citetalias{dflj18}. Although not mentioned by \citetalias{efe20}, they used $r_{\rm max}=50$ Mpc rather than $10$ Mpc for this analysis (H. Desmond, private communication).  Since both our estimate and theirs are based on $\Lambda$CDM simulations, this discrepancy is {\em not} an issue of $\Lambda$CDM versus modified gravity. Fig.~\ref{fig:aext-scaledep} shows that this also cannot be explained by the difference in $r_{\rm max}$; we see that the median $|\aext|$ increases by only a factor $\sim1.4$ from $r_{\rm max}=10$ Mpc to $50$ Mpc. Rather, it can be traced back to the combination of two effects: (i) the use of untruncated NFW (henceforth, uNFW) profiles by \citetalias{dflj18} in their estimate of \aext\ and (ii) the selection of the SPARC sample.

To understand the effect of not truncating, it is useful to pretend that all the mass of a halo is concentrated into a point at its center.  Then \aext\ is given by the first of the equalities in equation~(\ref{eq:aext-sim}), except that the sum is now over halos, so $m_{\rm part}$ is replaced by $m_h$ (different for each halo).  The question is what to use for $m_h$.  The mass within radius $r$ around an uNFW profile diverges logarithmically as $r\to\infty$. So, for the $i^{\rm th}$ neighbour with radius $R_i$ and concentration $c_i$ at separation $r_i$ from the galaxy in question, the assumption of an uNFW profile leads to a logarithmic enhancement of $\sim\ln(c_ir_i/R_i)/\ln(c_i)$ to the mass, and hence to the contribution of this neighbour to \aext.  For neighbours inside $10\,{\rm Mpc}$ with masses $\gtrsim10^{11}\Mh$ as used by \citetalias{dflj18}, $\ln(c_ir_i/R_i)/\ln(c_i)$ is typically a factor of $\sim4$, (and is about $\sim5$ when using $r_{\rm max}=50$ Mpc as in \citetalias{efe20}). 

Of course, because halos below some threshold will not be observed, truncating the profiles of the objects which are observed, at say their virial radius is guaranteed to \emph{underestimate} the actual mass distribution around the galaxy, since it neglects the mass in the environment between these neighbours. Thus, the uNFW assumption might coincidentally account for all mass reasonably well. To check, we repeated our analysis by replacing the sum over particles in \eqn{eq:aext-sim} with a mass-weighted sum over haloes having $m_{\rm vir}\geq7.7\times10^{10}\Mh$ \citepalias[close to the threshold used by][]{dflj18}, which gives us a point mass neighbour estimate for \aext. Multiplying this by a factor $4$ gives an accurate approximation to the uNFW neighbour estimate used by \citetalias{dflj18}. We find that the median $|\aext|/a_0$ of this uNFW estimate is $\sim9.0\times10^{-3}$, a factor $\sim1.5$ larger than our estimate using all the mass \emph{for the same mock galaxies}: The uNFW assumption does overestimate $|\aext|$. This overestimate is further modified in the \citetalias{dflj18} method because they add to the halo-based mass distribution the contribution of a smoothed reconstruction based on Lagrangian perturbation theory (see their section 2.3).

Regarding sample selection, we note that the typical value of $|\aext|/a_0$ inferred by \citetalias{dflj18} using the 2M++ sample \citep{lh11} is $\sim0.013$ (see fig. 3b of \citetalias{dflj18}), at least a factor $2$ smaller than that inferred from the SPARC sample and a factor $\sim1.4$ larger than the `uNFW neighbour' estimate using our mock catalog above. Since the SPARC and 2M++ analyses used essentially the same estimates of $\aext$ from \citetalias{dflj18}, their difference in typical $|\aext|/a_0$ (apart from a factor $\sim1.4$ due to the different $r_{\rm max}$ values, see above) must arise from differences in the underlying sample definitions (with SPARC being a biased subset of 2M++). Extending this reasoning to our mock catalog, we conclude that sample selection strategies make it difficult to compare our results for $|\aext|/a_0$ with the existing literature to within a factor of $2$.

A final interesting effect is that using the uNFW neighbour estimate as described above for our mock galaxies leads to a Spearman correlation coefficent between $\varepsilon\leftrightarrow|\aext|$ of $\sim+0.18$ with negligible $p$-value, which is twice as strong as the one inferred using the particle-based estimate of \aext. This can be traced back to the fact that the neighbour-based $|\aext|$ estimate correlates much more strongly with host mass $m_{\rm vir}$ than does the particle-based one (as \citealp{phs18} note, there is substantial scatter between $b_1$ and $m_{\rm vir}$). Thus, not only does the uNFW neighbour assumption overestimate the value of $|\aext|$, it also substantially overestimates the strength of the correlation between $|\aext|$ and $\varepsilon$. 

For now, we conclude that we have a fair understanding of the difference between our calculation of $|\aext|/a_0$ and the value inferred from the SPARC galaxies using the technique of \citetalias{dflj18}. As mentioned previously, as far as estimation techniques are concerned, our particle counting technique is the more reliable, since it correctly accounts for all the mass surrounding the galaxy in question. In the future, constrained simulations could be used to address the fidelity of methods which account for mass which is not observed (e.g., the technique of \citetalias{dflj18}), as well as the effects of sample selection, in more detail.  

\subsection{Correlation with $m_{\rm vir}$}
The results above indicate that the comparison between $e$ and $e_{\rm env}$ in \citetalias{efe20} is not robust. Therefore, we now focus on the $e\leftrightarrow m_{\rm vir}$ correlation, which is present in our $\Lambda$CDM mocks (c.f. Fig.~\ref{fig:aext-epsilon}, keeping in mind that our $\varepsilon$ behaves qualitatively like $-e$), and should not be affected by explicit environmental systematics.  For this, we need $m_{\rm vir}$ estimates for the SPARC sample.  We use the values from \citet{llms20} which were obtained by fitting a cored NFW profile with $\Lambda$CDM priors on baryon-dark matter scaling relations to SPARC rotation curves.  One caveat to be noted is that, for half of the sample of 148 objects, $\chi^2/{\rm dof}>2$, indicating a bad fit. (Some objects have an indeterminate $\chi^2/{\rm dof}$.) Fitting with an Einasto profile gave only slightly better results (see their Fig.~1). The corresponding error estimates on the $m_{\rm vir}$ values, on the other hand, are typically small ($\lesssim0.1$ dex) for most objects. This feature of the \citet{llms20} mass-modelling, namely, bad fits with small parameter errors, may indicate that the $m_{\rm vir}$ estimates are not reliable.  The formalism and results presented in \citetalias{pcs21} and \citetalias{ps21} suggest that a self-consistent CDM analysis of rotation curves, especially for RAR studies, \emph{must} include the effect of quasi-adiabatic relaxation, for example by allowing $q_{\rm rdm}$ (see section~\ref{subsec:mock}) to be a free parameter for each galaxy, in addition to halo mass, concentration and baryonic parameters.  This will allow a more reliable estimate of $m_{\rm vir}$.
We will take up the required Monte Carlo fitting exercise in the near future. For now, we proceed using $e$ estimates from \citetalias{efe20} and $m_{\rm vir}$ estimates from \citet{llms20}, along with their respective errors, without modification, but noting the caveats associated with point (2) at the start of this Section.

\begin{figure}
\centering
\includegraphics[width=0.475\textwidth]{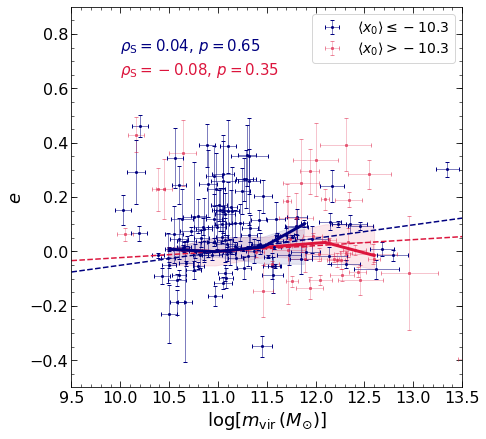}
\caption{{\bf SPARC $e$ against $\log[m_{\rm vir}]$.} Symbols show values and errors of $\log[m_{\rm vir}(M_\odot)]$ reported by \citet{llms20} and $e$ reported by \citet{efe20}, restricted to the 148 
galaxies used by \citet{efe20}. Blue points show the 113 objects having $\avg{x_0}<-10.3$ which were used for the EFE analysis by \citet[][see their fig.~5]{efe20}, while red points show the remaining objects.  Blue (red) label at the top shows the Spearman rank correlation coefficient $\rho_{\rm S}$ and associated $p$-value when using the blue symbols (all measurements) and ignoring errors. Dashed blue (red) line shows the result of orthogonal distance regression accounting for errors in both $e$ and $\log[m_{\rm vir}]$ for the blue symbols (all measurements). Thick solid blue (red) line and associated band shows the weighted median along with weighted central $68\%$ region of $e$ in bins of $\log[m_{\rm vir}]$ for the blue symbols (all measurements). The weights were taken to be proportional to the inverse variance of each measurement, and the location of the points on the horizontal axis was chosen to be the weighted median of $\log[m_{\rm vir}]$ in each bin.}
\label{fig:sparc-efe}
\end{figure}

Fig.~\ref{fig:sparc-efe} shows a scatter plot of $e$ against $\log[m_{\rm vir} (\Msun)]$ for this sample, with the points used (excluded) by \citetalias{efe20} shown in blue (red). We perform three statistical analyses to test for a correlation between these variables in the full sample as well as the low-acceleration subsample. The first is to ignore errors and calculate the Spearman rank correlation coefficient between $e$ and $m_{\rm vir}$. The values are reported in the labels in Fig.~\ref{fig:sparc-efe} and indicate no significant correlation. Next, we perform an orthogonal distance linear regression between $e$ and $\log[m_{\rm vir}]$, accounting for errors on both variables, using the \texttt{scipy.odr} numerical package. The results are shown as the red (blue) dashed line for all 148 (the $113$ low-acceleration) galaxies, with a regression slope of $0.022\pm0.008$ ($0.05\pm0.01$) treating $\log[m_{\rm vir}]$ as the independent variable. Finally, we calculate the inverse variance-weighted median and central $68\%$ region of $e$ in bins of $\log[m_{\rm vir}]$, shown as a function of the weighted median $\log[m_{\rm vir}]$ in each bin by the solid curves with error bands using the same colour coding; these agree quite well with the linear regression.

Overall, we conclude that, when accounting for errors in both variables, there is a weak but significant positive trend detected between $e$ and $m_{\rm vir}$ for the $113$ low-acceleration galaxies, which substantially weakens for the full sample. A positive correlation is the opposite of what is predicted by the CDM framework, as we discussed in  section~\ref{subsec:mockEFE} (there, \emph{lower} mass implies a downward deviation, or positive $e$), and instead appears consistent with the MOND expectation of a stronger decline in rotation curves (i.e., more positive $e$) in denser environments. However, the fact that the trend in the full sample is weaker than in the low-acceleration subsample, along with the caveat regarding the reliability of the $m_{\rm vir}$ estimates and the uncertainties on the errors of the $e$ values in the first place, suggests that this trend must be treated with caution. For now, we simply conclude that our results motivate a more self-consistent CDM analysis of the EFE in observed samples.

\section{Conclusions}
\label{sec:conclude}
We have presented a simple, exact analytical calculation showing that, in the general relativistic CDM framework, it is natural to expect a statistical correlation between the shape and/or amplitude of a galaxy's RAR and the strength of the external gravitational field \aext. This by no means implies a violation of the strong equivalence principle, but emerges instead from the clustered nature of the distribution of galaxies and their surrounding mass. 

We explicitly demonstrated this `external field effect' (EFE) in a $\Lambda$CDM-based mock catalog of rotation curves of massive spiral galaxies, showing that the amount by which a galaxy's RAR departs from the sample median ($\varepsilon$ of equation~\ref{eq:eps-def}) has a weak but significant positive correlation with its large-scale linear bias $b_1$ and the estimated $|\aext|$, simply because $\varepsilon$, $b_1$ and $\aext$ all correlate with the mass and concentration of the galaxy's host halo.  Although our $\varepsilon$ is not the same as the quantity $e$ which quantifies the {\em shape} of a galaxy's RAR, the fact that we see an $\varepsilon$-$|\aext|$ correlation strongly suggests that simply detecting a statistical EFE-like effect is {\em not} a conclusive test of GR.  In particular, our work suggests that it is important to compare any observed $e$-$|\aext|$ correlation with a similar measurement in $\Lambda$CDM simulations (e.g., $e$ should be obtained from treating the rotation curves and baryonic masses in $\Lambda$CDM simulations as though they were of real galaxies, so that the expected $e$-$|\aext|$ correlation is quantified).

On the other hand, the {\em sign} of the EFE-like correlation in our mocks may provide a useful test.  E.g., while MOND predicts that the $\varepsilon\leftrightarrow |\aext|$ correlation is negative, in $\Lambda$CDM the sign of this correlation depends on $m_{\rm vir}$: for stellar mass-selected samples, the sign is predicted to be positive. Moreover, in our mock catalog, the EFE-like $\varepsilon\leftrightarrow b_1$ and $\varepsilon\leftrightarrow |\aext|$ correlations vanish if measured at fixed mass and concentration. This vanishing may provide a better `null hypothesis' in the search for unexpected EFE-like effects.  Of course, in the $\Lambda$CDM context, the tightness of the RAR means that unexpected EFE-like effects may provide an efficient way to search for `assembly bias'.

Finally, we argued that recent claims \citepalias{efe20} of a statistical EFE detection in the SPARC sample should be treated with caution. This is partly due to the large uncertainties associated with extracting the EFE signal from fits to the RAR of individual galaxies. We have also found that the values of $|\aext|$ used by \citetalias{efe20}, which are determined following a $\Lambda$CDM-based Local Volume reconstruction from \citetalias{dflj18}, are likely overestimated by at least a factor of $\sim1.5$.
More importantly, approximations made in numerically estimating \aext\ (such as the use of untruncated NFW profiles by \citetalias{dflj18}) can spuriously enhance the correlation between $|\aext|$ and EFE residuals extracted from the RAR. This, and allowing $a_0$ to differ between objects when estimating $e$, must be accounted for in future studies of the EFE.\footnote{A recent analysis by \citet{efe21} has replaced the use of $\Lambda$CDM simulations for estimating $\aext$ with a baryon-painting approximation in the MOND context. This is difficult, as it requires large correction factors to account for baryons which may be present but are not observed directly.  The resulting median $e_{\rm env}$ is weakly correlated with $e$ for individual galaxies, but the large uncertainties mean that the correlation is also consistent with zero.} 

Our results not only motivate a more self-consistent treatment of mass-modelling of rotation curves within the CDM framework, properly accounting for the quasi-adiabatic relaxation of the dark matter in the presence of the baryons in a galaxy's host halo, but also call for more robust estimates of the external gravitational field at the locations of Local Volume galaxies. This will require techniques that can access small-scale spatial information, e.g., those based on Voronoi tessellations of the galaxy distribution \citep{pa20}; we will explore these in future work.

\section*{Acknowledgments}
We thank K.-H. Chae for motivating us to study the external field effect in the CDM context, both he and H. Desmond for useful correspondence and comments on an earlier draft, and R. Srianand for useful discussions.
The research of AP is supported by the Associateship Scheme of ICTP, Trieste and the Ramanujan Fellowship awarded by the Department of Science and Technology, Government of India.
This work made extensive use of the open source computing packages NumPy \citep{vanderwalt-numpy},\footnote{\url{http://www.numpy.org}} SciPy \citep{scipy},\footnote{\url{http://www.scipy.org}} Matplotlib \citep{hunter07_matplotlib}\footnote{\url{https://matplotlib.org/}} and Jupyter Notebook.\footnote{\url{https://jupyter.org}}

\section*{Data Availability}
The mock catalogs underlying this work will be made available upon reasonable request to the authors.

\bibliography{references}
 
\label{lastpage}
\end{document}